\documentclass{nseJournal}

\begin{document}

\title{Measurements of the $^{27}{\rm Al}(\alpha,n)$ Thick Target Yield \\ Near Threshold} 

\addAuthor{K. Br{and}enburg}{a}
\addAuthor{G. Hamad}{a}
\addAuthor{\correspondingAuthor{Z.~Meisel}}{a}
\correspondingEmail{meisel@ohio.edu}
\addAuthor{C.~R. Brune}{a}
\addAuthor{D.~E. Carter}{a}
\addAuthor{J. Derkin}{a}
\addAuthor{D.~C. Ingram}{a}
\addAuthor{Y. Jones-Alberty}{a}
\addAuthor{B. Kenady}{a}
\addAuthor{T.~N. Massey}{a}
\addAuthor{M. Saxena}{a}
\addAuthor{D. Soltesz}{a}
\addAuthor{S.~K. Subedi}{a}
\addAuthor{J. Warren}{a}

\addAffiliation{a}{Institute of Nuclear \& Particle Physics \\ Department of Physics \& Astronomy \\ Ohio University, Athens, Ohio 45701, USA}

\addKeyword{thick-target yield}
\addKeyword{neutron background}
\addKeyword{NIF diagnostic}

\titlePage

\begin{abstract}
We present results from direct measurements of the $^{27}{\rm Al}(\alpha,n)$ thick target yield from laboratory energies $E_{\alpha}\approx$ 3-5~MeV, performed with the HeBGB neutron detector at the Edwards Accelerator Laboratory. Our measurements have a small energy cadence in order to address discrepancies and sparseness of thick-target yield data sets existing for this energy region. We find general agreement with existing data sets, including yields derived from cross section data, while resolving a discrepancy between existing thick-target yield data sets for $E_{\alpha}\approx4-5$~MeV. However, for $E_{\alpha}<3.5$~MeV, our results are substantially lower than previous thick-target yield data and somewhat larger than than yields calculated from existing cross section data. Our data complete the energy-range needed for estimates of the $^{27}{\rm Al}(\alpha,n)$ contribution to neutrino and dark matter detector backgrounds and result in increased viability of $^{27}{\rm Al}(\alpha,n)$ as a tool for fusion plasma diagnostics, e.g. at the National Ignition Facility.
\end{abstract}

\section{Introduction}
$^{27}\rm{Al}(\alpha,n)$ is a significant neutron source from transuranic waste material~\cite{Gehr03}, could potentially be used for high energy density plasma diagnostics at the National Ignition Facility (NIF) or the International Thermonuclear Experimental Reactor (ITER)~\cite{Cerj18}, is a potential source of neutrons from some molten salt reactor components~\cite{Ouya13}, contributes to dark matter detector backgrounds~\cite{Mei09}, and plays a role in shock-driven nucleosynthesis in core collapse supernovae~\cite{Howa74,Sube20}. As such, it is important to have a precise characterization this reaction. For nuclear applications, the thick-target yield is of particular importance, as $\alpha$-particles emitted from $\alpha$-decaying nuclides will generally fully stop within nearby components.

Five cross section data sets exist for the $^{27}{\rm Al}(\alpha,n)$ cross section for $\alpha$ beam energies $E_{\alpha}<7$~MeV~\cite{Will60,Howa74,Flyn78,Holm86}, though the data of Ref.~\cite{Will60} have a strong contribution from background. The data sets of Refs.~\cite{Will60,Holm86} and one set from Ref.~\cite{Flyn78} have a small energy spacing that enables variations in the cross section to be observed. Two of the data sets in this energy range come from $^{27}{\rm Al}$ targets of moderate thickness~\cite{Howa74,Flyn78}, resulting in a smoother excitation function. In principle, each of these can be converted to thick-target yields by integrating the ratio of the cross section and stopping power. However, there are discrepancies between Ref.~\cite{Flyn78} and \cite{Holm86}. Ref.~\cite{Holm86} attribute this to carbon contamination issues of Ref.~\cite{Flyn78}. As such, there is a large difference in the yield depending on the adopted cross section data set~\cite{West22}. Additionally, there is an uncertainty contribution due to the adopted stopping-power. For instance, when Ref.~\cite{Vlas15} calculated thick-target yields from the cross section data of Refs.~\cite{Howa74,Flyn78}, they required an arbitrary normalization to agree with the directly measured thick-target yields of Ref.~\cite{West82}.

To date, two thick-target yield data sets exist for $^{27}{\rm Al}(\alpha,n)$~\cite{West82,Roug83}. Ref.~\cite{West82} employed a neutron counter, while Ref.~\cite{Roug83} performed their measurements using the activation technique. The data sets are in agreement for $E_{\alpha}>5$~MeV, though the data set of Ref.~\cite{Roug83} has rather sparse energy spacing over the entire energy range of interest, while the data set of Ref.~\cite{West82} has a single point for $E_{\alpha}<4$~MeV (to be compared to the $^{27}{\rm Al}(\alpha,n)$ reaction threshold of 3.034~MeV). The discrepancy between the two data sets can result in different estimates for neutron backgrounds resulting from $\alpha$-decaying nuclides in environments with fissile materials or actinide contaminants. The presence of only two data points with $E_{\alpha}<3.5$~MeV for Ref.~\cite{Roug83}, and none for Ref.~\cite{West82}, hinders the use of $^{27}{\rm Al}(\alpha,n)$ as a potential diagnostic reaction for fusion plasmas, e.g. at NIF~\cite{Cerj18}.

To remedy these issues, we performed measurements of $^{27}{\rm Al}(\alpha,n)$ thick-target yields with a small energy cadence for $E_{\alpha}<5$~MeV using the HeBGB neutron counter~\cite{Bran22} at the Edwards Accelerator Laboratory at Ohio Unviversity~\cite{Meis17}. In Section~\ref{sec:Measurement} we describe our measurement approach. We discuss the results in Section~\ref{sec:results} and provide concluding remarks in Section~\ref{sec:conclusion}.

\section{Measurement Details}
\label{sec:Measurement}

Measurements were performed at the Edwards Accelerator Laboratory at Ohio University~\cite{Meis17}. The full set-up is described in  detail in Ref.~\cite{Bran22}, though we cover the most relevant details here. A helium beam produced by an Alphatross ion source was accelerated using a 4.5~MV T-Type tandem pelletron to an energy between 3-5~MeV, departing the accelerator as $^{++}{\rm He}$. The beam was analyzed with an energy-calibrated 90$^{\circ}$ dipole magnet and slits with a gap of 0.254~cm, resulting in a beam energy uncertainty of 0.36\%. The beam passed through a 0.5~cm diameter gold-plated collimator located approximately 48~cm upstream of the target location and was impinged on a thick $^{27}{\rm Al}$ target located on a gold-plated ladder at the center of the HeBGB moderator. Typical beam intensities on-target were $\sim$50~nA. 

\begin{figure}[ht!]
\begin{center}
\includegraphics[width=0.5\columnwidth]{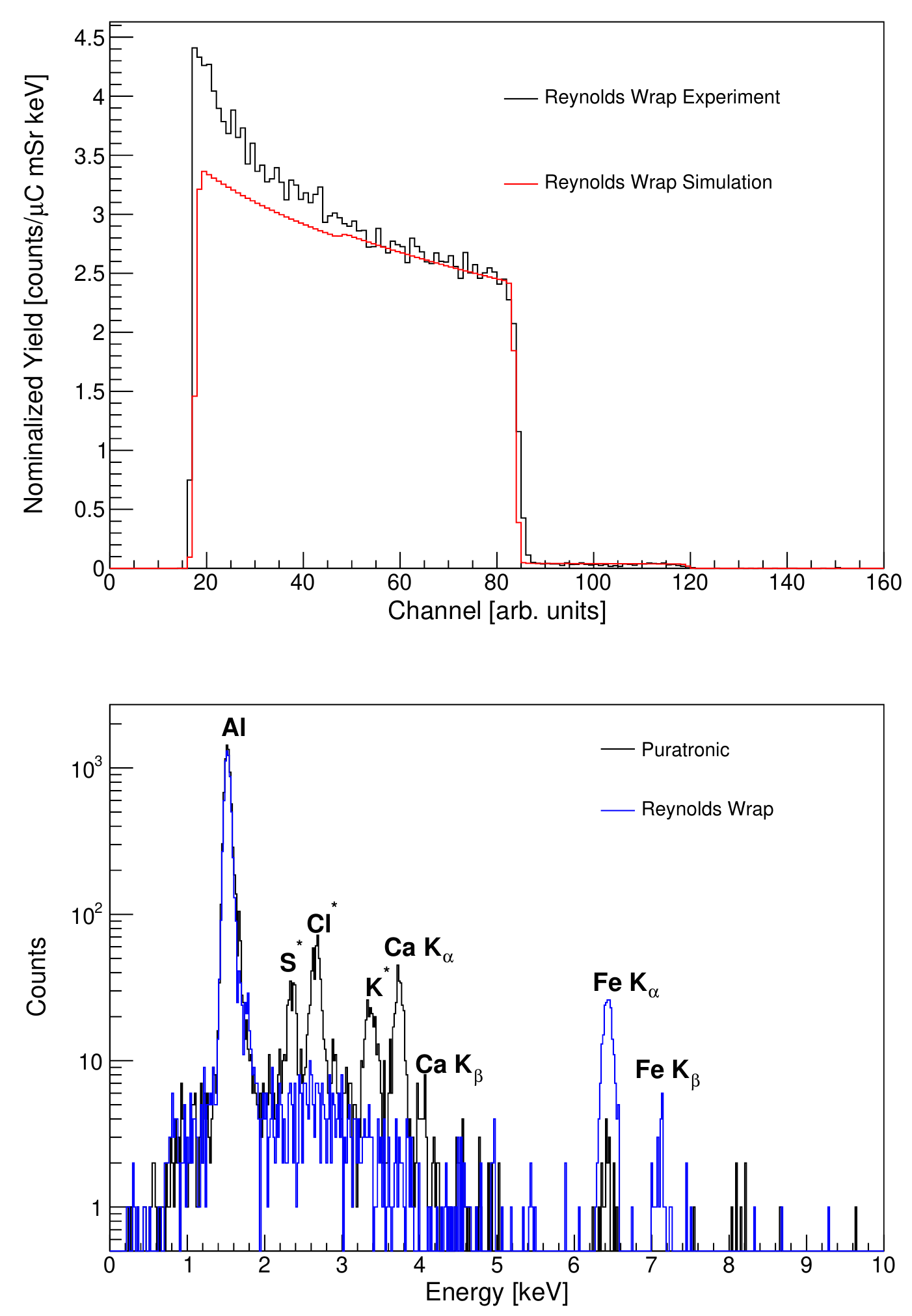}
\caption{Results from ion beam analysis of the aluminum target used in this work. The upper panel compares results of RBS measurements to calculations performed with {\tt RUMP}~\cite{Dool85}, where the main feature is due to aluminum and the small plateau from channels $\sim90-120$ is due to iron. The lower panel compares the results of PIXE measurements for our target material (Reynolds Wrap) to high-purity aluminum foil (Puratronic), where X-rays from elements present in the foils are identified. \label{fig:IBA}}
\end{center}
\end{figure}

The $^{27}{\rm Al}$ thick-target was made of Reynolds Wrap aluminum foil, cleaned by an alcohol solution, and folded-over several times to make a stopping target. Reynolds Wrap is manufactured to have high-purity of aluminum, where possible contaminants are $<$1\% by weight of magnesium, silicon, titanium, chromium, iron, manganese, copper, and zinc, and $<0.15$\% of all other elements~\cite{Ward95,Ward98}. We confirmed this purity using Rutherford backscattering (RBS) and particle-induced X-ray emission (PIXE) ion beam analysis, using the set-up described in Refs.~\cite{Meis17,Ingr17}. Results from RBS and PIXE measurements performed with a 2.2~MeV $\alpha$ beam are shown in Fig.~\ref{fig:IBA}. RBS results only show evidence for aluminum and iron, where calculations performed with {\tt RUMP}~\cite{Dool85} identified an iron abundance $<$1\% and confirmed that the abundance of the aforementioned suspected contaminants is $\ll$1\%. The deviation between the simulated and measured RBS spectra at low channel number (energy) is due to the absence of multiple-scattering effects in the {\tt RUMP} calculations~\cite{Baue93}. The PIXE spectrum in Fig.~\ref{fig:IBA} confirms that the target is primarily aluminum with some iron contamination and minor contributions from other elements. For context, a comparison is provided to Puratronic high-purity (99.99\% pure metals-basis) aluminum foil. 

For isotopes of the iron contamination identified in our target in Fig.~\ref{fig:IBA}, the $(\alpha,n)$ thresholds are 1.457~MeV ($^{57}{\rm Fe}$), 3.833~MeV ($^{58}{\rm Fe}$), 5.468~MeV ($^{56}{\rm Fe}$), and 6.255~MeV ($^{54}{\rm Fe}$). However, the Coulomb barrier suppresses $(\alpha,n)$ cross sections for these isotopes far below $^{27}{\rm Al}(\alpha,n)$ at the same $E_{\alpha}$. Consequently, thick-target yields for natural iron are around two-orders of magnitude (or more) lower than the thick-target yields from $^{27}{\rm Al}$~\cite{West82}. Additionally, we performed measurements of the neutron yield in HeBGB for $E_{\alpha}$ below the $^{27}{\rm Al}(\alpha,n)$ threshold and found yields on the order of the background (0.24$\pm$0.05~n/s for empty-target runs near $E_{\alpha}=3$~MeV). This provides further evidence that target contamination did not provide significant contributions to our measured $^{27}{\rm Al}(\alpha,n)$ thick-target yields for most energies. For energies just above threshold, the beam-on-target yields measured just below threshold were used to estimate the background contribution.

For our thick-target yield measurements, the incident charge was measured by summing the beam current of the target ladder and target chamber within HeBGB. To ensure the entirety of the beam was on-target, at each measurement energy the beam was first tuned through an empty space in the target ladder and the beam intensity $I_{\alpha}$ was recorded on a downstream Faraday cup. Empty target measurements were periodically recorded for enough time to provide high-statistics background measurements.  The number of detected neutrons $N_{n,{\rm det}}$ was counted using the $^{3}{\rm He}$ and ${\rm BF}_{3}$ neutron-sensitive proportional counters, configured as described in Ref.~\cite{Bran22}. A pulser, adjusted to provide a signal outside of the neutron spectrum, was used to monitor the proportional counter dead-time and therefore the live fraction $f_{\rm live}$ of the data acquisition.

While HeBGB provides nearly 4$\pi$ coverage, our {\tt MCNP} ~\cite{Goor12} simulation results show some dependence of the neutron detection efficiency $\epsilon$ on the neutron angular distribution~\cite{Bran22}. However, given the high excitation energy populated within the $^{31}{\rm P}$ compound nucleus by $^{27}{\rm Al}+\alpha$, a number of levels are likely to participate in the reaction and thus significantly anisotropic angular distributions are not anticipated. This is consistent with existing angular distribution data for this reaction~\cite{Batc59}. We therefore assume an isotropic angular distribution for our neutron detection efficiency.

The yield at each energy is calculated by $Y=N_{n,{\rm det}}/(I_{\alpha}t\epsilon f_{\rm live})$, where $I_{\alpha}t$ is the integrated charge on the target for a measurement and $\epsilon=7.5\pm1.2$\%. Our neutron detection efficiency dominates our yield uncertainty, with the exception of yields within a few-hundred keV of the reaction threshold, where the background uncertainty is significant. The integrated charge contributes a 1\% uncertainty. We add all uncertainties in quadrature to obtain our total yield uncertainty.

\begin{figure}[ht!]
\begin{center}
\includegraphics[width=0.45\columnwidth]{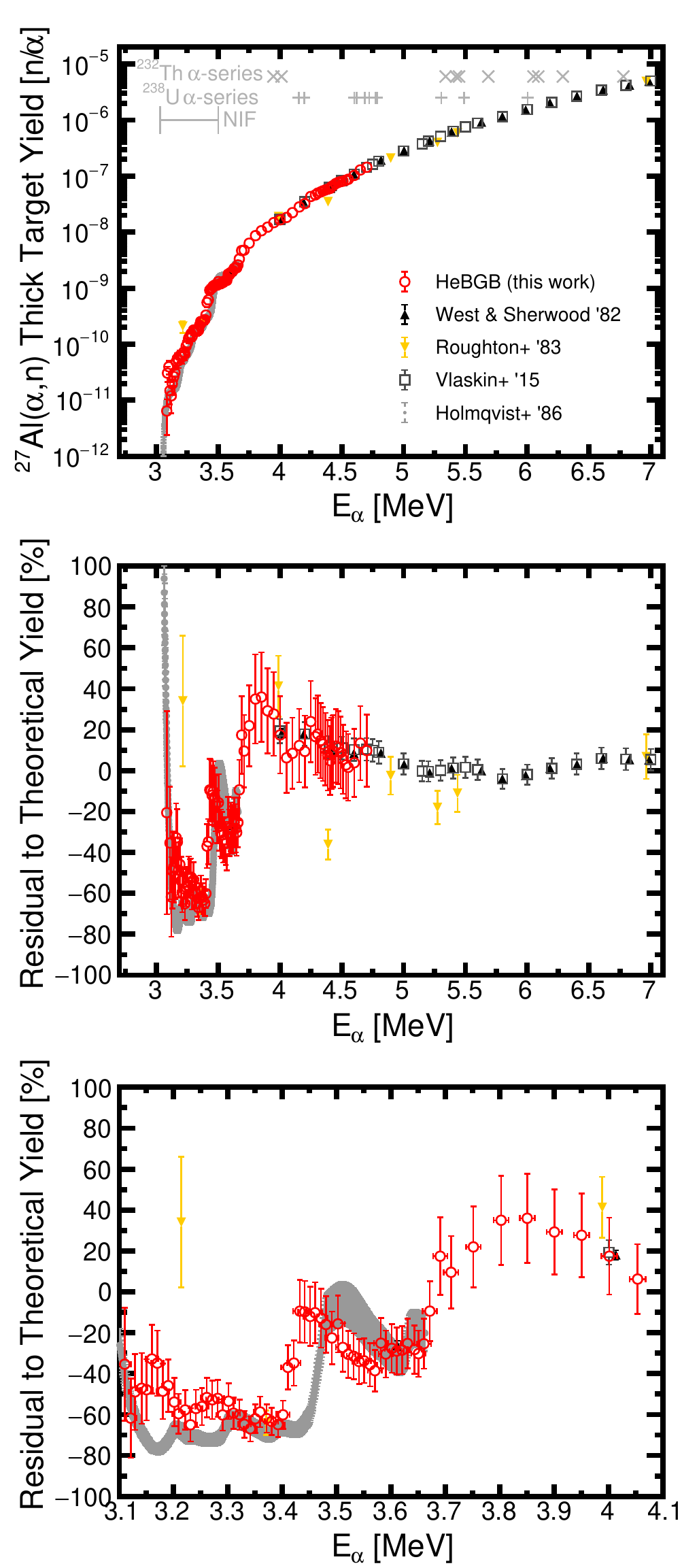}
\caption{The upper panel shows $^{27}{\rm Al}(\alpha,n)$ thick-target yields measured in this work (HeBGB), compared to yield measurements of Refs.~\cite{West82} (West \& Sherwood '82) and \cite{Roug83} (Roughton+ '83), the evaluated thick-target yield of Ref.~\cite{Vlas15} (Vlaskin+ '15), and the thick-target yield resulting from integrating the cross-section data of Ref.~\cite{Holm86} with stopping powers from Ref.~\cite{Zieg10}. The $^{232}{\rm Th}$ and $^{238}{\rm U}$ decay series $E_{\alpha}$ with decay branches greater than 1\% are also shown by the $\times$ and $+$, respectively, while a bracket $|$---$|$ indicates the energy range for $\alpha$-particles relevant for plasma fusion diagnostics (NIF). The middle panel shows a residual between the upper-panel results and theoretical yields calculated using the {\tt Talysv1.8}~\cite{Koni08} default cross section and {\tt SRIM2013}~\cite{Zieg10} stopping power, where the theoretical yields are arbitrarily scaled by a factor of 0.8. The lower panel is the same as the middle panel, but focusing on a narrower energy range.\label{fig:TTY}}
\end{center}
\end{figure}

\section{Results and Discussion}
\label{sec:results}

We present our thick-target yield measurement results\footnote{See the Supplemental Material for a table of our measured yield.} in the upper-panel of Fig.~\ref{fig:TTY} (also ), where we compare to thick-target yield measurement data of Ref.~\cite{West82,Roug83} and the evaluation results of Ref.~\cite{Vlas15}, who integrated the cross section data of Refs.~\cite{Howa74,Flyn78} with the stopping power of Ref.~\cite{Zieg10} and scaled the results to match Ref.~\cite{West82}. We also compare to thick-target yields calculated using thin-target cross section data of Ref.~\cite{Holm86} and the stopping power of Ref.~\cite{Zieg10}, where we include a 5\% uncertainty in the stopping power based on the world data for helium stopping in aluminum at these energies~\cite{Mont17}. For context, the upper panel of Fig.~\ref{fig:TTY} highlights $E_{\alpha}$ of typical actinide contaminants in neutrino and dark matter detectors~\cite{Gand11}, as well as the $E_{\alpha}$ range relevant for fusion plasma diagnostics~\cite{Cerj18}. In order to compare these thick-target yields in more detail, the middle and lower panels of Fig.~\ref{fig:TTY} shows the residual to an arbitrary theoretical yield. We calculated the $^{27}{\rm Al}(\alpha,n)$ cross section using default settings of {\tt Talysv1.8}~\cite{Koni08}, determined the stopping power using the stopping power from {\tt SRIM2013}~\cite{Zieg10}, and adjusted the resulting thick-target yield by an arbitrary normalization of $\times$0.8 in order to approximately go through the average trend of the data sets.

Our $^{27}{\rm Al}(\alpha,n)$ thick-target yields are overall in good agreement with existing data sets, where we agree with Ref.~\cite{West82} rather than Ref.~\cite{Roug83} at the energies where these two data sets disagree. Just above the reaction threshold, our results show slightly larger yields than yields calculated with the cross section data of Ref.~\cite{Holm86}, while being substantially lower than the lowest-energy thick-target yield data point of Ref.~\cite{Roug83}. The discrepancy near threshold is likely explained by background subtraction issues. Ref.~\cite{Holm86} performed background measurements for various $E_{\alpha}$ impinged on a gold backing and on their aluminum target while below threshold. It is possible that the background for the measurements performed with the gold backing was not the same as the background for the aluminum thick-target yield measurements. This could be due to different rates of carbon build-up, e.g. due to different quality of vacuum or different a beam tune impinging on more or less carbon-coated components upstream of the target location, or different room background rates due to changing environment conditions~\cite{Wals17}. For the results reported here, our target ladder system, including an empty frame for background measurements, enabled us to obtain frequent background measurements for the same $E_{\alpha}$, same beam tune, and same setup at almost the same time as the thick-target yield measurements.

Meanwhile, we also disagree with yields derived from Ref.~\cite{Holm86} cross sections as to the location of the step-like yield increase in the $E_{\alpha}=3.4-3.5$ energy range. While yields resulting from Ref.~\cite{Holm86} have a step-like increase (from a resonance-like feature in the cross section) at $E_{\alpha}=3.5$~MeV, we find that the increase occurs roughly 50~keV lower. This is likely due to the fact that Ref.~\cite{Holm86} actually measured thick-target yields, which they did not report, but rather differentiated to arrive their reported cross sections. Small uncertainties in the beam energy, as well as insufficiently fine energy spacing between measurement points, can compound to shift the energy of features in the resulting thick-target yield. Shifts could also occur due to adopting different stopping powers, but these uncertainties (discussed above) are included in the uncertainty band.

Our results demonstrate that the $^{27}{\rm Al}(\alpha,n)$ thick-target yields of Ref.~\cite{West82} are preferred over the results of Ref.~\cite{Roug83}. We complement the former data set, by extending $^{27}{\rm Al}(\alpha,n)$ thick-target yield data down to 60~keV above the reaction threshold. Our data complete the energy-range needed for estimates of the $^{27}{\rm Al}(\alpha,n)$ contribution to neutrino and dark matter detector backgrounds. Meanwhile, our results modify the $^{27}{\rm Al}(\alpha,n)$ yield that would be estimated for a fusion plasma diagnostic scenario, e.g. at NIF, relative to estimates based on prior thick-target data or thick-target yields calculated from existing cross section data. Specifically, our results imply larger yields from $^{27}{\rm Al}(\alpha,n)$, and therefore higher-precision constraints on the mix of ablator material in the the DT fuel~\cite{Cerj18}. Detailed calculations are highly complex, e.g. see Ref.~\cite{Lona22} for calculations of $^{10}{\rm B}(\alpha,n)$ yields at NIF, and are well beyond the scope of the present work.

\section{Conclusion}
\label{sec:conclusion}
We performed thick-target yield measurements of $^{27}{\rm Al}(\alpha,n)$ for $E_{\alpha}=3-5$ using the HeBGB neutron counter at the Edwards Accelerator Laboratory at Ohio University. Our results remedy the sparseness in thick-target yield data for $E_{\alpha}<5$~MeV and resolve the discrepancy between prior results of Ref.~\cite{West82} and \cite{Roug83}. Our directly measured thick-target yields are slightly larger than thick-target yields inferred from the cross section data of Ref.~\cite{Holm86}, who provide the only high-energy resolution cross section data near threshold, while being substantially lower than the only thick-target yield data reported in this energy range~\cite{Roug83}. Meanwhile, we find that the step-like increase in the yield near $E_{\alpha}\sim3.5$~MeV occurs $\sim$50~keV lower in energy relative to prior results~\cite{Holm86}. Our data complete the energy-range needed for estimates of the $^{27}{\rm Al}(\alpha,n)$ contribution to neutrino and dark matter detector backgrounds and result in increased viability of $^{27}{\rm Al}(\alpha,n)$ as a tool for fusion plasma diagnostics.
 
 \pagebreak
 
\section*{Acknowledgments}
This work was supported in part by the U.S. Department of Energy Office of Science under Grants No. DE-FG02-88ER40387 and DE-SC0019042 and the U.S. National Nuclear Security Administration through Grants No. DE-NA0003883 and DE-NA0003909. The helium ion source was provided by Grant No. PHY-1827893 from the U.S. National Science Foundation. We also benefited from support by the U.S. National Science Foundation under Grant No. PHY-1430152 (Joint Institute for Nuclear Astrophysics -- Center for the Evolution of the Elements).

\section*{Disclosure Statement}

The authors report there are no competing interests to declare.

\section*{Data Availability Statement}
Thick target yields reported in this work are available as a text file in the Supplemental Material.

\pagebreak
\bibliographystyle{bst/ans_js}                                                                           
\bibliography{References}

\end{document}